\documentclass[runningheads]{llncs}
\pdfoutput=1
\usepackage{graphicx}
\usepackage{amssymb}
\usepackage{amsmath}
\usepackage{breqn}
\usepackage[colorlinks]{hyperref}
\newcommand{\myfootnote}[1]{
    \renewcommand{\thefootnote}{}
    \footnotetext{\scriptsize#1}
    \renewcommand{\thefootnote}{\arabic{footnote}}
}

\usepackage{booktabs}       
\usepackage{tabularx}

\usepackage{multirow}

\begin{document}
\title{Degenerative Adversarial NeuroImage Nets:\\ Generating Images that Mimic Disease Progression}
\titlerunning{Degenerative Adversarial NeuroImage Net}
%
%

\author{Daniele Ravi \and Daniel C. Alexander \and Neil P. Oxtoby \and for the Alzheimer's Disease Neuroimaging Initiative*}


%
\authorrunning{Daniele Ravi et al.}

\institute{Centre for Medical Image Computing (CMIC), \\Department of Computer Science, University College London\\
\email{d.ravi@ucl.ac.uk}\\}

\maketitle
\begin{abstract}
Simulating images representative of neurodegenerative diseases is important for predicting patient outcomes and for validation of computational models of disease progression. This capability is valuable for secondary prevention clinical trials where outcomes and screening criteria involve neuroimaging. Traditional computational methods are limited by imposing a parametric model for atrophy and are extremely resource-demanding. Recent advances in deep learning have yielded data-driven models for longitudinal studies (e.g., face ageing) that are capable of generating synthetic images in real-time. Similar solutions can be used to model trajectories of atrophy in the brain, although new challenges need to be addressed to ensure accurate disease progression modelling. Here we propose Degenerative Adversarial NeuroImage Net (DaniNet) --- a new deep learning approach that learns to emulate the effect of neurodegeneration on MRI {\color{black}by simulating atrophy as a function of ages, and disease progression. DaniNet uses an underlying set of Support Vector Regressors (SVRs) trained to capture the patterns of regional intensity changes that accompany disease progression. DaniNet produces whole output images, consisting of 2D-MRI slices that are constrained to match regional predictions from the SVRs. DaniNet is also able to maintain the unique brain morphology of individuals}. Adversarial training ensures realistic brain images and smooth temporal progression. We train our model using 9652 T1-weighted (longitudinal) MRI extracted from the Alzheimer's Disease Neuroimaging Initiative (ADNI) dataset. We perform quantitative and qualitative evaluations on a separate test set of 1283 images (also from ADNI) demonstrating the ability of DaniNet to produce accurate and convincing synthetic images that emulate disease progression.

\end{abstract}
\myfootnote{*Data used in preparation of this article were obtained from the Alzheimer's Disease Neuroimaging Initiative
(ADNI) database (adni.loni.usc.edu). As such, the investigators within the ADNI contributed to the design
and implementation of ADNI and/or provided data but did not participate in analysis or writing of this report.
A complete listing of ADNI investigators can be found at:
http://adni.loni.usc.edu/wp-content/uploads/how_to_apply/ADNI_Acknowledgement_List.pdf}
\section{Introduction}
Neurodegenerative diseases are strongly age-related, and the incidence of these diseases is expected to rise in the next few years due to increasing life expectancy. Creating tools that assist our limited understanding of neurodegeneration can help to answer some of the open questions in this field, e.g., What are the causes? How can we improve subtype classification? 

Disease progression modelling~\cite{oxtoby2017imaging} analyzes clinical and image-based biomarkers to map out longitudinal changes during chronic diseases. However, modelling temporal neurodegeneration on full resolution MRI is still a major challenge. To tackle this problem, a few simulators have been proposed in the literature~\cite{camara2006phenomenological,karaccali2006simulation,sharma2010evaluation,modat2014simulating}. The recent SimulAtrophy~\cite{khanal2017simulating} uses a computational model based on fluid mechanics. This approach combines two deformation fields: one from a biophysical model and the other obtained by non-rigid registration of two real images. For a given baseline MRI, the combined deformation field is used to impose the desired level of atrophy and generate the simulated image. SimulAtrophy is extremely resource-demanding and is not scalable to high-resolution images. Exploiting the success of deep learning, Bowles et al.~\cite{bowles2018modelling} proposed a framework based on Generative Adversarial Networks (GANs) to model and manipulate MRI directly through the technique of image arithmetic. Although the system is able to introduce or remove atrophy patterns from regions in the brain, it presents a few important limitations: disease progression is modelled linearly and morphological changes are the same across all patients. A more advanced deep learning framework developed to predict face ageing was proposed in~\cite{zhang2017age}. Our idea is to use a similar framework to generate realistic images that also preserve biological/anatomical constraints associated with disease progression. 

In this paper we present a new framework that overcomes the limitations of existing approaches. {\color{black}Firstly, similar to~\cite{zhang2017age} we employ adversarial training to ensure a high level of realism in the synthetic MRI. Secondly, we avoid imposing predefined atrophy patterns, prefering to implement novel biological constraints that model neurodegeneration. These constraints control the MRI intensity in each brain region through a set of localised SVRs learnt directly from the data. Finally, we design DaniNet to handle non-imaging characteristics, which we employ here to condition upon age and diagnosis}. Experiments evaluate the performance of the various components of DaniNet and demonstrate the ability to produce accurate and realistic synthetic images that emulate disease progression.

\section{Methods}
\begin{figure*}[t!]
 \begin{center}
 \includegraphics[width=0.85\textwidth,trim={0cm 5.6cm 15.9cm 0cm},clip]{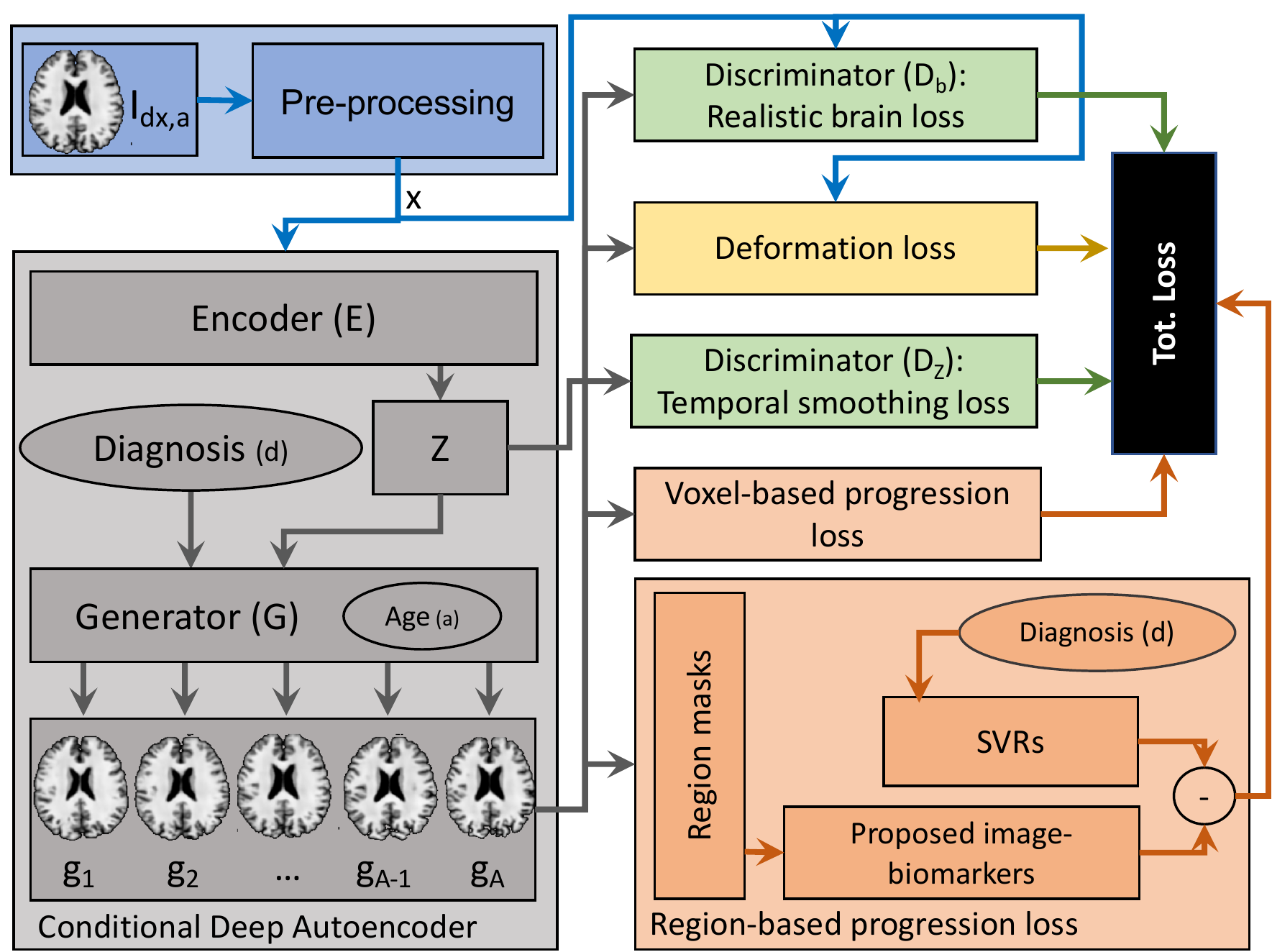}
 \caption{Pipeline used for training the proposed DaniNet framework. Each component of the pipeline is identified by a different colour.}
 \label{fig:DaniNet} 
\end{center}
\end{figure*}

In this section we present the main blocks which compose DaniNet. A colour-coded work-flow of our system is depicted in Fig.~\ref{fig:DaniNet}.

The first pre-processing block (shown in blue in Fig.~\ref{fig:DaniNet}) extracts a normalized slice $x$ from the MRI $I$. The second component (shown in grey) is a Conditional Deep Autoencoder (CDA) composed of two deep neural networks: an encoder $E$ that embeds $x$ in a latent space $Z$, and a generator $G$ that projects the vectors produced by $E$ back to the original manifold. Before this projection, the latent vector $z$ is conditioned with two variables: i) $d$ --- a numerical representation {\color{black} [0-3]} of diagnosis (i.e. cognitively normal, subjective memory concern, early/late mild cognitive impairment, Alzheimer's disease); and ii) $a \in [1,A]$ --- an index describing age, binned into {\color{black}$A$=10} groups. This age discretisation is important for computing deformation loss (block in yellow) to learn morphological changes along the progression. The third component (shown in green), consists of two discriminator networks: i) $D_z$ that drives $E$ to produce $z$ with a uniform prior and smooth temporal progression; and ii) $D_{b}$ that drives $G$ to produce realistic \textit{brain} neuroimages. Finally, in the blocks coded in orange, we present the proposed biological constraints used to model disease progression. In the next sub-sections, we will describe each of these components in detail.

\subsection{Pre-processing}
The pre-processing block removes irrelevant variation on the data and ensures that intensity values in each voxel only decrease. In an ideal scenario, intensity of T1-weighted MRI decreases with age since tissue density (having high intensity) will reduce, while water content (having low intensity) will increase. However, due to scanner variability and other sources of noise, this may not be the case in practice. To handle these variabilities, each input image is normalised using the following pre-processing steps: i) linear co-registration to a 1mm isotropic MNI template {\color{black}using FLIRT-FSL, ii) skull-stripping using BET-FSL}, and iii) intensity normalisation to zero mean and unit standard deviation. Images where pre-processing failed were not included in the training set.

\subsection{Conditional Deep Autoencoder}
A CDA is an extension of the Deep Autoencoder (DA) with the ability to integrate conditional variables inside the generator $G$. The advantage of training a single end-to-end CDA over training separate DAs is that it avoids overfitting individual groups when longitudinal data are missing. For the architecture of $E$ and $G$ we follow~\cite{zhang2017age}. The output of the encoder $E$ is a feature vector $z \in \mathbb{R}^s$ that preserves the brain morphology of the current participant. The output of $G$ are synthetic images defined as $g_a=G(E(x),a,d)$. The aim of $G$ is to learn the mapping between the linear transition of the latent vector conditioned on $a$ and $d$ and the non-linear transition in the original manifold.

\begin{figure*}[t!]
 \begin{center}
 \includegraphics[width=0.85\textwidth,trim={0cm 5.2cm 14.6cm 0cm},clip]{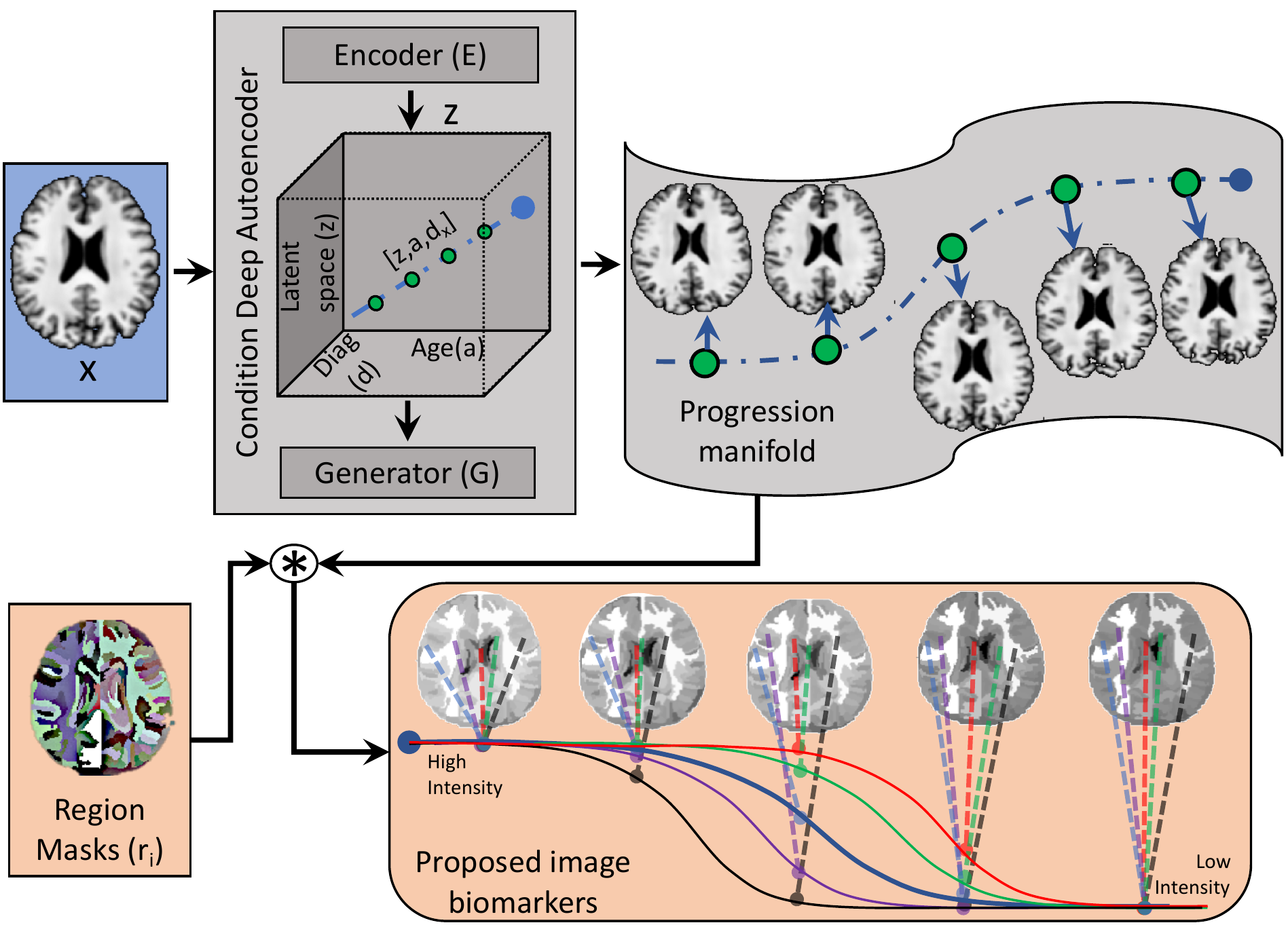}
 \caption{Pipeline used to obtain the proposed image biomarkers based on regional-based intensity progression.}
 \label{fig:Biomarker} 
\end{center}
\end{figure*}

\subsection{Adversarial Training}
Again following~\cite{zhang2017age}, DaniNet includes two discriminator networks that are trained adversarially with the CDA. The first discriminator $D_z$ guides $E$ to generate $z$ with a uniform distribution $\mathbb{U}$ to ensure temporal smoothness, as demonstrated in~\cite{zhang2017age}. Specifically, $D_z$ is trained to distinguish the vector generated by the encoder $E$ from samples extracted from $\mathbb{U}$. On the other side, $E$ is trained in a zero-sum game with the purpose to fool $D_z$. The objective function used for this adversarial training is

\begin{equation}\label{dz_loss}
\min\limits_{E}\max\limits_{D_z}\mathbb{E}_{z^*}\big[ \log D_{z}(z^*) \big]+\mathbb{E}_{x}\big[1- \log D_{z}(E(x)) \big],
\end{equation}
where $\mathbb{E}$ is the expectation operator, $z^*$ is a vector sampled from $\mathbb{U}$, $D_{z}$ estimates the probability that a vector comes from $\mathbb{U}$, and $E(x)$ is the latent vector obtained from $x$. 

The second discriminator $D_{b}$ guides the generator $G$ to produce realistic brain images. This discriminator is trained to distinguish the synthetic images $g_a$ from the real images in the training set via the following objective function

\begin{equation}\label{dbrain_loss}
\min\limits_{G}\max\limits_{D_{b}}\mathbb{E}_{x}\big[\log D_{b}\big(x)\big) \big] + 
\mathbb{E}_{x}\big[1-D_{b}\big(G(E(x),a,d)\big)\big],
\end{equation}
where $D_{b}$ estimates the probability that a slice is extracted from a real MRI.

\subsection{Biological Constraints}
To capture the patterns of image intensity changes that accompany disease progression, our framework uses two separate loss functions during training. These loss functions impose biological constraints that mimic neurodegeneration by ensuring decreased intensity (tissue density) that is consistent with disease progression.

The first loss function operates at the voxel level. Given a synthetic output $g_a$ in age group $a$, we impose that corresponding voxels of each preceding image $g_i$ with $i<a$ have higher-intensity values. Likewise, later images $g_j$ with $j>a$ have lower-intensity values. We express this constraint as
\begin{equation}\label{L_vox}
L_{\textrm{vox}}=\frac{1}{MN(A-1)}\bigg[\displaystyle\sum_{p=1}^{a-1}\textrm{sgn}(g_a-g_p)+
\displaystyle\sum_{p=a+1}^{A}\textrm{sgn}(g_p-g_a)\bigg], 
\end{equation}
where $M$ and $N$ are the number of rows and columns in each slice{\color{black}, and sgn is the sign function}. 

The second loss function operates at a regional level to improve spatial consistency across neighbouring voxels. We pre-train a set of SVRs to learn how to predict the rate of intensity progression in fixed, overlapping, data-driven regions mask obtained from hierarchical clustering of local intensity of the training images. On the considered axial slice the clustering algorithm detects 128 regions with an average size of 233 voxels. We train one SVR per region, where the input features are the age at baseline, the age at the follow-up, and diagnosis $d$. We restrict the SVR training to monotonically decreasing data by removing time-points where regional intensity increases. The loss function used to constrain the image sequence $g_i$ (with $i \in [1,A]$) through the regional SVR predictions is

\begin{multline}\label{L_reg}
L_{\textrm{reg}}=\frac{1}{R(A-1)}\displaystyle\sum_{i=1}^{R}\bigg[\displaystyle\sum_{p=1}^{a-1}\bigg(\textrm{SVR}_i(p,a,d)-\frac{\sum[g_a\cdot r_i]+\epsilon}{\sum[g_p\cdot r_i]+\epsilon}\bigg)+
\\
\displaystyle\sum_{p=a+1}^{A}\bigg(\textrm{SVR}_i(a,p,d)-\frac{\sum[g_p\cdot r_i]+\epsilon}{\sum[g_a\cdot r_i]+\epsilon}\bigg)\bigg],
\end{multline}
where $R$ is the number of regions, $r_i$ is the i-th region-mask, $SVR_i(p,a,d)$ is the prediction of intensity rate progression for the region $i$, and $\epsilon=0.1$ avoids division by 0. 

The pre-trained SVRs supply a set of constraints on the generator's outputs for ensuring accurate emulation of neurodegeneration. The process used to compute the rate of regional intensity progression is summarized in Fig.~\ref{fig:Biomarker}.

\subsection{Deformation Loss}
Our final loss function ensures consistency with the individual subject over time by minimising the difference between individual input image sequences and the corresponding output sequences. For each input image $x$, the difference is minimised between $x$ and a weighted average of two outputs from the nearest age bins:

\begin{equation}\label{def}
L_{\textrm{def}}=L_{2}\bigg(x , \big[g_a\alpha+g_{a+1}(1-\alpha)\big]\bigg),
\end{equation}
where $\alpha$ reflects the distance between input age and the group $a$ and $a+1$. 

In summary, the loss functions $L_{\textrm{reg}}$ and $L_{\textrm{vox}}$ ensure monotonic intensity change to mimic neurodegeneration, and $L_{\textrm{def}}$ together with $D_{b}$ ensure realistic brain morphology.

\subsection{Training Details, Parameters and Evaluation}
Data used in the preparation of this article were obtained from the ADNI database (adni.loni.usc.edu). The ADNI was launched in 2003 as a public-private partnership, led by Principal Investigator Michael W. Weiner, MD. The primary goal of ADNI has been to test whether serial magnetic resonance imaging (MRI), positron emission tomography (PET), other biological markers, and clinical and neuropsychological assessment can be combined to measure the progression of mild cognitive impairment (MCI) and early Alzheimer's disease (AD).

Training data ({\color{black}MRI-slices: 9852; participants: 876) and testing (MRI-slices: 1283; participants: 179) are pre-processed T1-weighted MRI from the ADNI dataset. This includes data from multiple sites, scanners, and pre-processing pipelines. We included them all to improve robustness to such variability, and hence increase generalizability. The participants were aged between 63 and 87 years old, with the following distribution of clinical disease stage: 28\% cognitively normal, 4\% subjective memory concern, 54\% mild cognitive impairment and 14\% Alzheimer's disease.} Each participant has on average 4.7 MRI spanning 3 years. We select participants in the test set having at least one follow-up visit two years after baseline, to allow sufficient time for observable neurodegeneration to occur.

{\color{black}
The architectures of each network $E$, $G$, $D_{b}$, $D_Z$ are based on the implementation proposed in~\cite{zhang2017age}. The size of the latent space $Z$ is fixed to 200. The network's parameters of $E$ and $G$ are trained to minimize a total loss defined as
\begin{equation}\label{Tot}
L_{\textrm{tot}} = E_z * 0.0003 + G_b * 0.0003 + L_{\textrm{vox}} * 0.0003 +L_{\textrm{reg}} * 0.0003 + L_{\textrm{def}} * 0.2
\end{equation}
where $E_z$ is the cross entropy obtained by the discriminator $D_z$ on the generated latent vectors, and $G_b$ is the cross entropy obtained by the disctiminator $D_b$ on synthetic images $g_a$. To train our system we use the stochastic gradient descent solver, ADAM ($\alpha$ = 0.0002, $\beta$1 = 0.5) on an NVIDIA GTX TITAN-X GPU card. The training procedure converged after 3000 epochs of random mini-batch of 100 slices with a size of 128$\times$128. 
We also align each test individual to the model by baseline age and diagnosis, then personalise DaniNet with a transfer learning step ($T$ component, see results section~\ref{resultRection}) that involves an additional 200 training iterations on the single baseline image. Personalisation is essential to tune the model for the specific morphology of the individual's brain, ready to simulate realistic, personalised followup MRI. To note, only a single MRI from each test subject's first visit is used to personalise the trained model.}

In our evaluation, input $x$ are from baseline visits. Follow-up data is the ground truth $y$ against which we evaluate generated images $g_i$ using complementary quantitative and qualitative analysis. The quantitative analysis involves the use of the Structural Similarity Index matrix (SSIM). Since this matrix is not able to quantify the actual realism of the obtained images we also perform a complementary qualitative analysis via a survey of clinicians and medical imaging experts that evaluate the synthetic images based on their perception.

\begin{table*}[t]
\scriptsize
\caption{Performance of DaniNet on different configurations.}
\label{table:Configuration}
\begin{center}
\resizebox{0.90\textwidth}{!}{%
\begin{tabular}{c|>{\centering\arraybackslash}p{1cm}>{\centering\arraybackslash}p{1cm}>{\centering\arraybackslash}p{1cm}>{\centering\arraybackslash}p{1cm}>{\centering\arraybackslash}p{1cm}>{\centering\arraybackslash}p{1cm}>{\centering\arraybackslash}p{1cm}>{\centering\arraybackslash}p{1cm}}
 \multirow{ 4}{*}{\rotatebox{90}{SSIM}}& ~\cite{zhang2017age} & $C$ & $T$ & $C$-$T$ &$P$ &$P$-$C$ &$P$-$T$ & $P$-$C$-$T$\\
\cline{2-9}
&0.43 $\pm$ 0.07&0.44 $\pm$ 0.06&0.59 $\pm$ 0.14&0.60 $\pm$ 0.14&0.44 $\pm$ 0.08&0.48 $\pm$ 0.08&0.61 $\pm$ 0.15&\textbf{0.62 $\pm$ 0.16} \\

\end{tabular}
}
\end{center}
\end{table*}
\normalsize

\section{Results}\label{resultRection}
Our first experiment is designed to show that DaniNet improves similarity between generated image and ground truth, with respect to the baseline approach developed for face ageing~\cite{zhang2017age}. Specifically, we evaluate the contribution obtained by three components of DaniNet: i) $P$ obtained exploiting $L_{\textrm{vox}}$ and $L_{\textrm{reg}}$, ii) $C$ obtained conditioning the system with the clinical diagnosis, and iii) $T$ obtained when the learned model is transferred before testing on new data.

\begin{figure*}[t!]
 \begin{center}
 \includegraphics[width=1\textwidth,trim={0cm 3.1cm 1.7cm 0cm},clip]{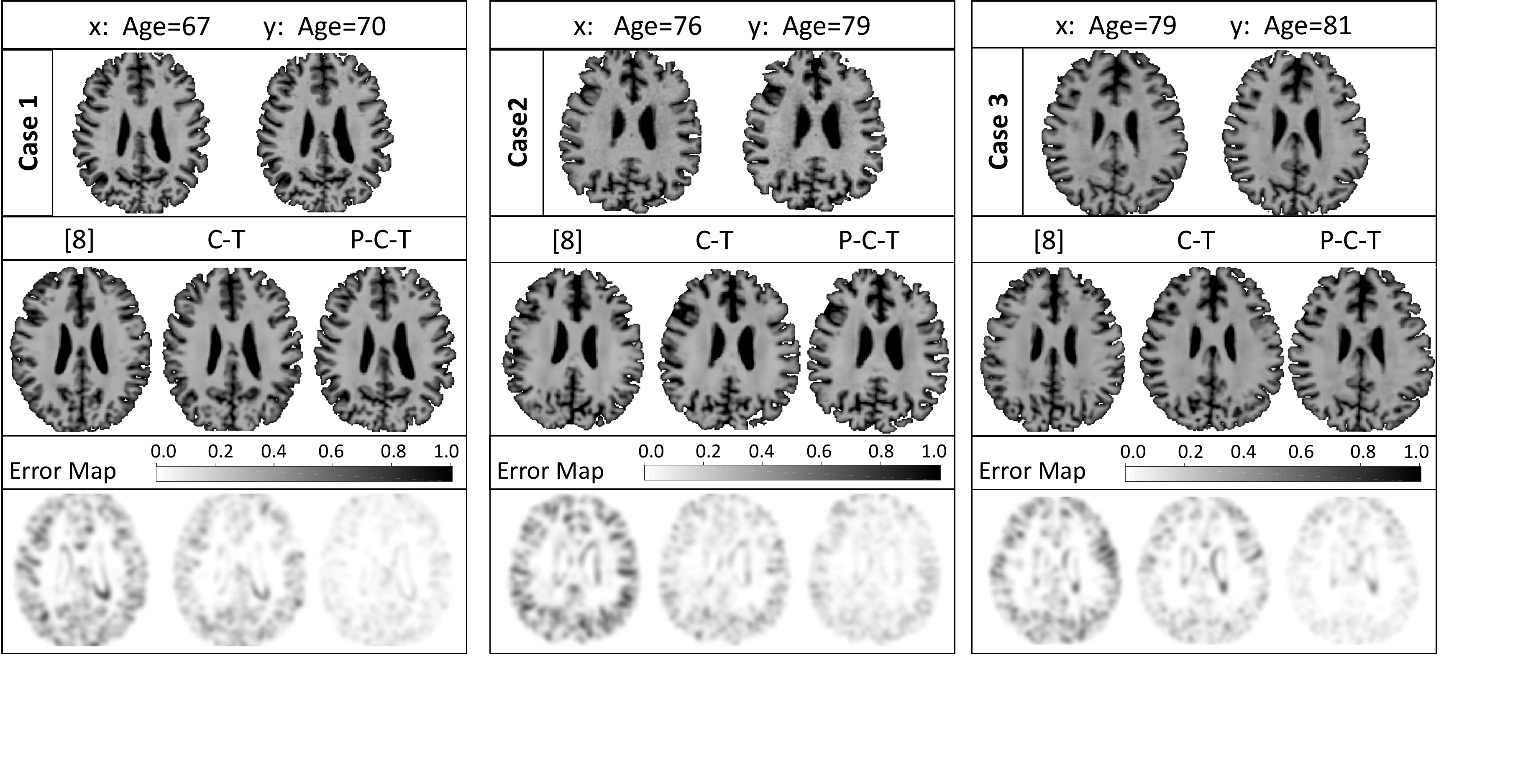}
 \caption{Visual results obtained by different configurations of DaniNet on MRI slices from three participants in the test set.}
 \label{fig:ResultGTComparison} 
\end{center}
\end{figure*}

\begin{figure*}[t!]
 \begin{center}
 \includegraphics[width=1\textwidth,trim={0cm 12.3cm 8.5cm 0cm},clip]{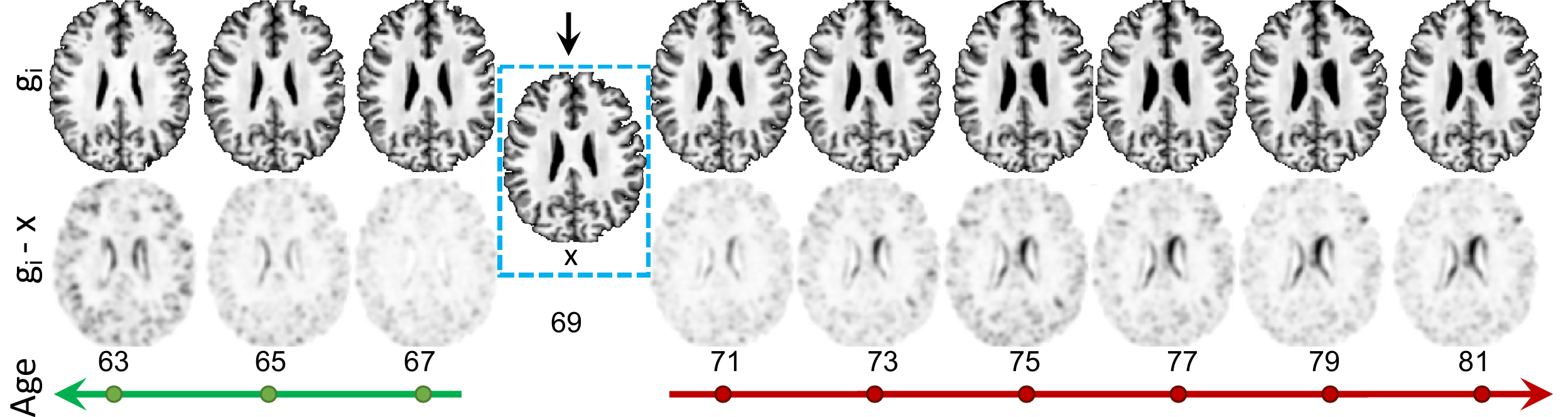}
 \caption{Neurodegeneration simulation of a 69-year old ADNI participant.}
 \label{fig:progression} 
\end{center}
\end{figure*}

For this purpose, the SSIM results obtained using different compositions of $P$, $C$, and $T$ are reported in Table~\ref{table:Configuration}. Clearly, the best results are obtained when the framework includes all three components. The baseline model produces inferior SSIM. Transfer learning $T$ is the feature that provides the highest contribution. A moderate contribution is provided by $P$ and $C$. These improvements were assessed statistically using a paired t-test and all p-values were less than 0.0001. Visual results for the configurations $C$-$T$, $P$-$C$-$T$ and baseline~\cite{zhang2017age} are shown in Fig.~\ref{fig:ResultGTComparison}. The error maps, shown in the last row of the figure, confirm our findings. A full simulation with the optimal configuration $P$-$C$-$T$ is shown in Fig.~\ref{fig:progression} where neurodegeneration is apparent in the progression (left to right), most obviously in ventricular expansion and cortical thinning. 

Our second experiment tests the qualitative performance of DaniNet based on human visual perception. In a survey, we asked 25 medical imaging experts (neurologists and computer scientists) to evaluate on average 36 sets of three images each, randomly selected from the test set, generated using two different configurations of DaniNet ($C$-$T$ and $P$-$C$-$T$) and the baseline~\cite{zhang2017age}. These were shown to the user in a random order. The input $x$ and the related follow-up $y$ were also displayed on the screen as references for the participants. For each set of images, the user was asked to select the closest synthetic image to $y$. Results from the survey confirm that DaniNet is a considerable improvement of the baseline approach. The medical imaging experts selected the configuration $P$-$C$-$T$ $26\pm5$ (mean $\pm$ std) times ($72\pm14\%$), the configuration $C$-$T$ $7\pm4$ times ($19\pm11\%$), and the baseline $1\pm1$ times ($3\pm3\%$). For only $2\pm3$ outputs ($6\pm8\%$) the users were not happy with any of the generated synthetic images.

\section{Conclusion and Future Work}
We have proposed and evaluated (quantitatively and qualitatively) a novel deep-learning framework that is able to learn how to emulate the effect of neurodegenerative disease progression on structural MRI. {\color{black} The framework produces personalised, realistic output images through a combination of biological constraints, transfer learning, and conditioning upon both fixed and variable non-imaging characteristics. To the best of our knowledge, we are the first to propose a simulator that imitates realistic neurodegeneration by imposing biological constraints. In future work we will extend the framework to simulate entire 3D-MRI with the aid of low-memory techniques~\cite{blumberg2018DIQT}. Additionally, focal brain pathologies, such as glioma and white matter hyperintensities, are not currently modelled by our framework; future work will extend to consider such effects.} Finally, although we have demonstrated our framework for modelling atrophy in MRI, we believe that this solution can be used with different image modalities (e.g. PET, CT, etc.) and to model disease progression on other organs (e.g lung, retina, etc.).

\section*{Acknowledgement}
The authors would like to thank NVIDIA Corporation  for the donation of the Titan Xp GPU used for this research. 

Data collection and sharing for this project was funded by the ADNI  (National Institutes of Health Grant U01 AG024904) and DOD ADNI (Department of Defense award number W81XWH-12-2-0012). ADNI is funded by the National Institute on Aging, the National Institute of Biomedical Imaging and Bioengineering, and through generous contributions from the following: AbbVie, Alzheimer's Association; Alzheimer's Drug Discovery Foundation; Araclon Biotech; BioClinica, Inc.; Biogen; Bristol-Myers Squibb Company; CereSpir, Inc.; Cogstate; Eisai Inc.; Elan Pharmaceuticals, Inc.; Eli Lilly and Company; EuroImmun; F. Hoffmann-La Roche Ltd and its affiliated company Genentech, Inc.; Fujirebio; GE Healthcare; IXICO Ltd.; Janssen Alzheimer Immunotherapy Research \& Development, LLC.; Johnson \& Johnson Pharmaceutical Research \& Development LLC.; Lumosity; Lundbeck; Merck \& Co., Inc.; Meso Scale Diagnostics, LLC.; NeuroRx Research; Neurotrack Technologies; Novartis Pharmaceuticals Corporation; Pfizer Inc.; Piramal Imaging; Servier; Takeda Pharmaceutical Company; and Transition Therapeutics. The Canadian Institutes of Health Research is providing funds to support ADNI clinical sites in Canada. Private sector contributions are facilitated by the Foundation for the National Institutes of Health (www.fnih.org). The grantee organization is the Northern California Institute for Research and Education, and the study is coordinated by the Alzheimer's Therapeutic Research Institute at the University of Southern California. ADNI data are disseminated by the Laboratory for Neuro Imaging at the University of Southern California. 

This project has received funding from the European Union's Horizon 2020 research and innovation programme under grant agreement No. 666992.

EPSRC grant EP/M020533/1 supports DCA's work on this topic. The NIHR UCLH Biomedical Research Centre also supports this work.

\bibliographystyle{splncs04}


\end{document}